\documentclass{PoS}

\usepackage{amsmath} 

\title{Scalar QED\(_2\) with a topological term -- \\ a lattice study in a dual representation}

\ShortTitle{Scalar QED\(_2\) with a topological term}

\author{\speaker{Thomas Kloiber}, Christof Gattringer\\
        Institut f\"ur Physik, Karl-Franzens-Universit\"at, 8010 Graz, Austria \\
        \email{thomas.kloiber@uni-graz.at} \\ \email{christof.gattringer@uni-graz.at}}

\abstract{We present a dual representation for the partition function of 2-dimensional scalar quantum electrodynamics with a topological term 
(\(\theta\)-term). In the dual representation the complex action problem at non-zero \(\theta\) is absent, which is an obstacle for Monte Carlo simulations in the conventional form of the model. We discuss the technical aspects of the dual representation and show that a dual Monte Carlo simulation can be implemented. As a first application we demonstrate how the \(2\pi\)-periodicity of physical observables is recovered in a suitable continuum limit.}

\FullConference{The 32nd International Symposium on Lattice Field Theory\\
		 23-28 June, 2014\\
		 Columbia University New York, NY}

\begin{document}

\section{Introduction}

When a topological term (\(\theta\)-term) is coupled to QCD (and some other theories), the action has a non-vanishing imaginary part. 
The Boltzmann factor can no longer be interpreted as a probability distribution and the theory is not directly accessible with importance sampling 
methods. From a technical point of view this ``complex action problem'' is similar to the one encountered in systems with non-zero chemical 
potential. Recently the complex action problem has been solved for some systems by rewriting their partition sum in terms of dual variables
where the partition sum has only real and positive terms and a Monte Carlo simulation is possible (see, e.g., \cite{dual}). 
 
Here we apply the dual approach to scalar QED\(_2\) with a topological term, i.e., the scalar Schwinger model. We show that also a quartic 
self-interaction of the scalars can be included, such that the theory can also be viewed as the abelian gauge Higgs model in two dimensions.
We show that the lattice version of this model can be mapped to a dual representation where also for finite vacuum angle $\theta$ the partition sum has only real and positive contributions. In terms of the dual variables a Monte Carlo simulation becomes possible and the main part of this contribution is dedicated to the technical aspects of the dualization and the algorithm.  

Since the \(\theta\)-parameter couples to a topological charge which is integer valued in the continuum limit, all observables should be \(2\pi\)-periodic in \(\theta\) in that limit. As a first physical result we show that this is indeed the case when taking a properly defined continuum limit.

The model under consideration has been studied from various points of view, e.~g., \cite{abelianhiggs} and also in higher dimensions \cite{abelianhiggsdim}. However, in these studies the complex action problem was overcome only partially, restricting the accessible region of the \(\theta\) angle to a small interval or even to zero. A study related to our future aims can be found in \cite{GIMP87}, where the behavior of topological objects (vortices) in different phases analogous to type-I and type-II superconductors was investigated. This, however, was a classical analysis and to our knowledge there have not been any simulations of the quantized theory. Some rigorous results for this model can be found in \cite{JT80}.

\section{Formulation of the model}
\label{sec_model}

In the conventional representation the Euclidean continuum action for the scalar Schwinger model with a quartic self interaction 
(U(1) gauge-Higgs model in 2-d) is given by
\begin{equation}
S  \; = \; \int d^{\,2}x ~\Big[ (D_\mu \phi^*)(D_\mu \phi)+m^2\phi^*\phi+\lambda(\phi^*\phi)^2 + \frac{\beta}{4} F_{\mu\nu}F_{\mu\nu} \Big]~,
\end{equation}
with the covariant derivative $D_\mu = \partial_\mu + i A_\mu$,
the field strength tensor
$F_{\mu\nu} = \partial_\mu A_\nu - \partial_\nu A_\mu$, the mass parameter $m$, the self-interaction coupling 
of the matter fields $\lambda$ and the inverse gauge coupling $\beta$.

The  topological term ($\theta$-term) involves the topological charge $Q$, which is given by
\begin{equation}
\label{eq_top_charge}
Q \; = \; \frac{1}{4\pi} \int d^2x ~ \epsilon_{\mu\nu} F_{\mu\nu}  \; = \; \frac{1}{2\pi} \int d^2x ~F_{12} \; .
\end{equation}
The partition sum of the model is $Z = \int \mathcal{D}[A]~\mathcal{D}[\phi]~e^{-\, S \, -i\,\theta \, Q}$. 
Obviously the Boltzmann factor is complex for $\theta \neq 0$ and the theory has a complex action problem.

For the lattice simulations we have to define the model on a two dimensional  $N_s \times N_t$ space-time lattice $\Lambda$ with periodic 
boundary conditions for all fields. We split the action $S$ into gauge and matter parts, i.e., 
$S[U,\phi] = S_G[U] + S_M[U,\phi]$. 
For the gauge part we use the Wilson gauge action
\begin{equation}
\label{eq_conv_gauge}
S_G[U]=-\frac{\beta}{2} \sum_{x\in \Lambda} \big[U^p_x+U^{p\; *}_x\big]~.
\end{equation}
The plaquette variables $U^p_x$ are given by the usual
product of the link variables $U_{x,\nu} \in $ U(1), i.e.,  
$U^p_x = U_{x,1} U_{x+\hat{1},2} U_{x+\hat{2},1}^*U_{x,2}^*$. 
The matter part can be discretized as ($\kappa \equiv 4 + m^2$)
\begin{equation}
S_M[U,\phi]=\sum_{x \in \Lambda} \Big[\kappa |\phi_x|^2+\lambda|\phi_x|^4-\sum_\nu \big(\phi^*_x U_{x,\nu} \phi_{x+\hat{\nu}} + \phi_x U_{x,\nu}^* \phi^*_{x+\hat{\nu}}\big) \Big]~.
\end{equation}
 
There exist various approaches for the discretization of the topological charge $Q$ on the lattice. For the dualization we explore here, the so-called 
field theoretical definition turns out to be the best choice, i.e., a direct discretization of the continuum charge (\ref{eq_top_charge}). Writing the 
link variables as $U_{x,\nu} = e^{i A_\nu(x)}$ (the lattice spacing is set to $a = 1$ throughout this paper), the plaquette reads 
$U^p_x=e^{i\, (A_1(x)+A_2(x+\hat{1})-A_1(x+\hat{2})-A_2(x))}$ and after expansion in small $A_\nu$ one finds  
$U^p_x= 1 + i F_{12}(x) + {\mathcal O}(A^2)$.
A suitable combination of $U^p_x$ and $U^{p\;*}_x$ then yields the field theoretical definition of the topological charge on the lattice,
\begin{equation}
\label{eq_conv_topcharge}
Q[U] = \frac{1}{i 4 \pi} \sum_x \big[U^p_x -U^{p\; *}_x \big]~.
\end{equation}
Putting things together we arrive at the following form of the partition function
\begin{equation}
\label{eq_conv_part}
Z  \; = \; \int \!\! \mathcal{D}[U]\mathcal{D}[\phi]~e^{-\, S_G[U] \, - \, S_M[U,\phi] \, -i\,\theta \, Q[U]} \; = \; 
\int \!\! \mathcal{D}[U]\mathcal{D}[\phi]~e^{-\, S_M[U,\phi] } \; e^{ \, \eta \sum_x U^p_x} \; e^{ \, \overline\eta \sum_x U^{p\;*}_x} \; ,
\end{equation}
with 
\begin{equation}
\eta = \frac{\beta}{2}-\frac{\theta}{4\pi}\; , \; \; \overline{\eta} = \frac{\beta}{2}+\frac{\theta}{4\pi}
\quad \mbox{and} \quad 
\int \!\! \mathcal{D}[U] \mathcal{D}[\phi] \equiv \prod_x ~\int_\mathbb{C} d\phi_x ~~\prod_{x,\nu} ~\int_{U(1)} dU_{x,\nu}~.
\end{equation}
It is obvious, that there is a complex action problem in Eq. (\ref{eq_conv_part}) for non-zero values of the \(\theta\) angle 
(then $\eta \neq \overline\eta\,$), i.e., the Boltzmann factor acquires a phase and is therefore not suitable as a probability weight for importance sampling.

\section{Dual Representation}
\label{sec_dual}

\noindent
We can write the partition function as 
\begin{equation}
Z  = \int \! \mathcal{D}[U]  \; \prod_x e^{ \, \eta  U^p_x} \; \prod_x e^{ \, \overline\eta U^{p\; *}_x} \; Z_M[U]  \qquad, \qquad Z_M[U]  = \int  \mathcal{D}[\phi]~e^{- \, S_M[U,\phi]} ~ ,
\end{equation}
where \(Z_M[U]\) is the partition function of the matter fields for a given gauge configuration. This partition sum can be rewritten to dual variables 
along the way outlined in \cite{U1dual}. $Z_M[U]$ assumes the form of a sum over closed loops that are represented by integer link variables
$l_{x,\nu} \in \mathbb{Z}$, $\bar{l}_{x,\nu} \in \mathbb{N}_0$ which are subjects to constraints (see below). 
The configurations of these link variables (i.e., of the loops) are weighted with real and positive weight factors (also given below). The loops are 
dressed with link variables $U_{x,\nu}$ along their contour. 
 
In a second step (compare again \cite{U1dual}) also the exponentials $e^{ \, \eta U^p_x}$ and $e^{ \, \overline{ \eta }U^{p\;*}_x}$ are expanded 
using integer valued expansion indices  
assigned to the plaquettes of the lattice. The link variables $U_{x,\nu}$ from the plaquettes, together with the link variables 
from the loops in $Z_M[U]$ can then be integrated out. This leads to new constraints that connect the integer valued link variables $l_{x,\nu}$ with
the expansion indices for the plaquette terms $e^{ \, \eta U^p_x}$ and $e^{ \, \overline{ \eta }U^{p\;*}_x}$.
Thus, after a straightforward modification of the derivation in \cite{U1dual}, the partition sum (\ref{eq_conv_part}) 
is exactly rewritten to its dual form:
\begin{equation}
\label{eq_dual_part}
\begin{split}
Z = \sum_{\{l,\bar{l},p,\bar{p}\}} ~ &\prod_{x,\nu} ~\frac{1}{(|l_{x,\nu}|+\bar{l}_{x,\nu})! \bar{l}_{x,\nu}!} ~ \prod_x P(n_x)\\
&\prod_x ~\frac{\eta^{(|p_x|+p_x)/2+\bar{p}_x} ~ \overline{\eta}^{(|p_x|+p_x)/2-\bar{p}_x}}{(|p_x|+\bar{p}_x)!~\bar{p}_x!}\\
& \prod_x~ \delta(p_x-p_{x-\hat{2}}+l_{x,1})~\delta(p_{x-\hat{1}}-p_x+l_{x,2})~ \delta \Big(\sum_\nu \big[l_{x,\nu}-l_{x-\hat{\nu},\nu}\big]\Big)~.
\end{split}
\end{equation}
We have introduced 
\begin{equation}
P(n_x) \equiv \int_0^\infty dr ~ r^{n_x+1} ~ e^{-\kappa r^2 - \lambda r^4}~, ~~~~~ n_x \equiv \sum_\nu \Big[ |l_{x,\nu}|+|l_{x-\hat{\nu},\nu}|
+2(\bar{l}_{x,\nu}+\bar{l}_{x-\hat{\nu},\nu})\Big]~.
\end{equation}
The new degrees of freedom
\begin{equation}
l_{x,\nu},~p_x \in \mathbb{Z}~, ~~~~~ \bar{l}_{x,\nu},~\bar{p}(x) \in \mathbb{N}_0~,
\end{equation}
are constrained due to the Kronecker deltas, coming from the phase factor integration. The \(p\)- and \(\bar{p}\)-fields are associated with the gauge 
fields and live on the plaquettes of the lattices, hence called plaquette 
variables,  and the \(l\)- and \(\bar{l}\)-fields come from the matter fields and are attached to 
the lattice links and therefore are called link variables.

We observe, that in the dual representation \(Z\) has only real and positive terms, as long as  
\begin{equation}
\label{eq_pos_cond}
\beta > \frac{\theta}{2\pi} \; ,
\end{equation}
i.e.,  as long as \(\eta\) is positive, and Monte Carlo studies with importance sampling are feasible in the dual representation. The condition 
(\ref{eq_pos_cond})
does not impose a problem in practice, because one is interested in the continuum limit, i.e., \(\beta\rightarrow\infty\) and thus 
arbitrary values of \(\theta\) can be simulated.

\section{Update algorithm for the dual simulation}

Due to the constraints imposed on the dual variables, only certain admissible 
configurations contribute in (\ref{eq_dual_part}). However, the \(\bar{l}\)- and \(\bar{p}\)-fields do not appear in the 
constraints and thus can be updated independently 
in the usual way with a local Metropolis update scheme.
The \(p\)- and \(l\)-fields mix in the constraints and a strategy to update them jointly is necessary. Here we use the following
local update scheme of the variables attached to a site $x$:
We choose \(\Delta=\pm 1\) with equal probability and offer a trial configuration
$l^\prime, p^\prime$ as follows:
\begin{equation}
\begin{gathered}
p_x \rightarrow p'_x = p_x + \Delta~,\\
l_{x,1} \rightarrow l'_{x,1}=l_{x,1}-\Delta~,~~~l_{x+\hat{1},2} \rightarrow l'_{x+\hat{1},2}=l_{x+\hat{1},2} -\Delta~,\\
l_{x+\hat{2},1} \rightarrow l'_{x+\hat{2},1}=l_{x+\hat{2},1}+\Delta~,~~~l_{x,2} \rightarrow l'_{x,2}=l_{x,2}+\Delta~,
\end{gathered}
\label{update1}
\end{equation}
and all other $l$- and $p$ variables remain unchanged.
As can easily be checked also the trial configuration $l^\prime, p^\prime$ obeys all the constraints (if the original configuration $l,p$ did).
Furthermore every admissible configuration can be reached with the proposed changes, i.e., the update is ergodic. The trial configuration is then 
accepted or rejected with a Metropolis step. 

However, it turned out, that due to the close connection of the topological charge 
to configurations with constant plaquette values, $p_x = n, \forall x$,
(see below), a special pure gauge update reduces auto-correlation times considerably. Therefore we mix the update 
(\ref{update1}) with another update, which involves only the dual gauge fields \(p\): Again we choose \(\Delta=\pm1\) with equal probability and 
update the \(p\)-fields at all sites
\begin{equation}
p_x \rightarrow p'_x = p_x + \Delta ~~~~ \forall x~,
\end{equation}
i.e., we cover the whole \(2\)-torus with a sheet of \(p\) variables, all with the same value. Also this trial configuration is accepted or rejected with a 
Metropolis step. We remark, that it would also be straightforward to adopt the surface worm algorithm of \cite{U1dual} to the model studied here.

The algorithm has been thoroughly tested in the limit \(\theta = 0\) where we can use a Monte Carlo simulation without 
complex action problem in the conventional representation of the model. We also tested the algorithm in the case where we neglect the 
matter fields, i.e., the pure gauge case. Here one can obtain exact analytical results by explicitly summing up (\ref{eq_dual_part}). The only
admissible configurations in the pure gauge case are those with $p_x = n, \forall x$ and the partition sum can be expressed as a sum 
over modified Bessel functions. In both test cases we found perfect agreement 
within error bars, and, although not systematically studied, the dual simulations always outperformed the conventional ones.
This suggests the interpretation that the dual degrees of freedom provide the physically more natural representation of the system.

\section{Numerical Results}
\label{sec_results}

Bulk observables, i.e.,  various derivatives of the logarithm of the partition function with respect to the parameters of the model, are most easily 
accessible and have a very simple form in terms of the dual variables. 
In this exploratory study we focus on the expectation value of the plaquette
\begin{equation}
\langle \text{Re} (\square )\rangle \equiv \frac{1}{N_s~N_t} \frac{\partial \ln Z}{\partial \beta}=\frac{1}{2~N_s~N_t} \left[ \left\langle \frac{|\mathfrak{P}|+
\mathfrak{P}}{2} + \bar{\mathfrak{P}} \right\rangle \eta^{-1} + \left\langle \frac{|\mathfrak{P}|-\mathfrak{P}}{2} + \bar{\mathfrak{P}} \right\rangle 
\bar{\eta}^{-1} \right]~,
\end{equation}
and the topological charge
\begin{equation}
\langle Q \rangle \equiv \frac{1}{N_s~N_t} \frac{\partial \ln Z}{\partial \theta}=\frac{1}{2~N_s~N_t} \left[ \left\langle \frac{|\mathfrak{P}|+\mathfrak{P}}{2} + \bar{\mathfrak{P}} \right\rangle \eta^{-1} - \left\langle \frac{|\mathfrak{P}|-\mathfrak{P}}{2} + \bar{\mathfrak{P}} \right\rangle \bar{\eta}^{-1} \right]~,
\end{equation}
where we have defined  
\begin{equation}
|\mathfrak{P}| \equiv \sum_x |p_x|~,~~~~~\mathfrak{P} \equiv \sum_x p_x~~~~~\text{and}~~~~~\bar{\mathfrak{P}} \equiv \sum_x \bar{p}_x~.
\end{equation}
 
Here we study the observables as a function of \(\theta\) for different values of the mass parameter \(\kappa = 4\), \(5\) and \(10\), which 
corresponds to a vanishing bare mass, a relatively small $m$ and a large one. All these parameters lead to the unbroken Coulomb phase (we set the self-coupling to \(\lambda=1\)). 
The inverse gauge coupling $\beta$ is studied in an increasing sequence in order to approach the continuum limit,
since a main point of this first series of simulations is to establish \(2\pi\)-periodicity in \(\theta\) since for the field theoretical 
definition of \(Q\) which we use, \(2\pi\)-periodicity can emerge only in the continuum limit.  Dimensional analysis suggests that the continuum- 
and thermodynamical limits have to be taken in the following 
way
\begin{equation}
\beta \rightarrow \infty~, ~~~~~ |N_s~N_t| \rightarrow \infty ~~~~~ \text{with} ~~~~~ \frac{\beta}{N_s ~ N_t} = \text{const.} 
\end{equation}

\vspace{-0.5cm}

\begin{figure}[h!]
\begin{center}
\includegraphics[width=7.516cm,type=pdf,ext=.pdf,read=.pdf]{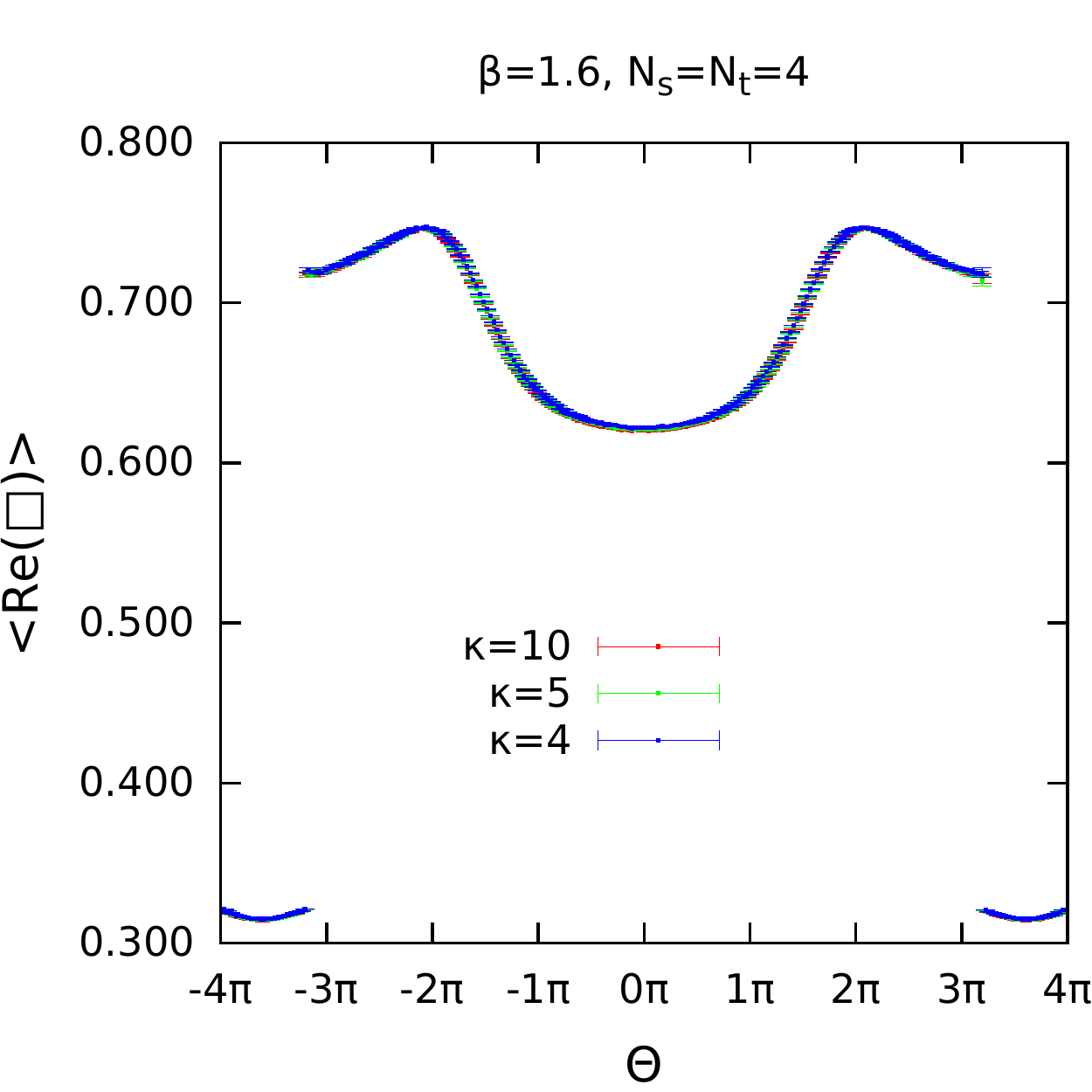} 
\includegraphics[width=7.516cm,type=pdf,ext=.pdf,read=.pdf]{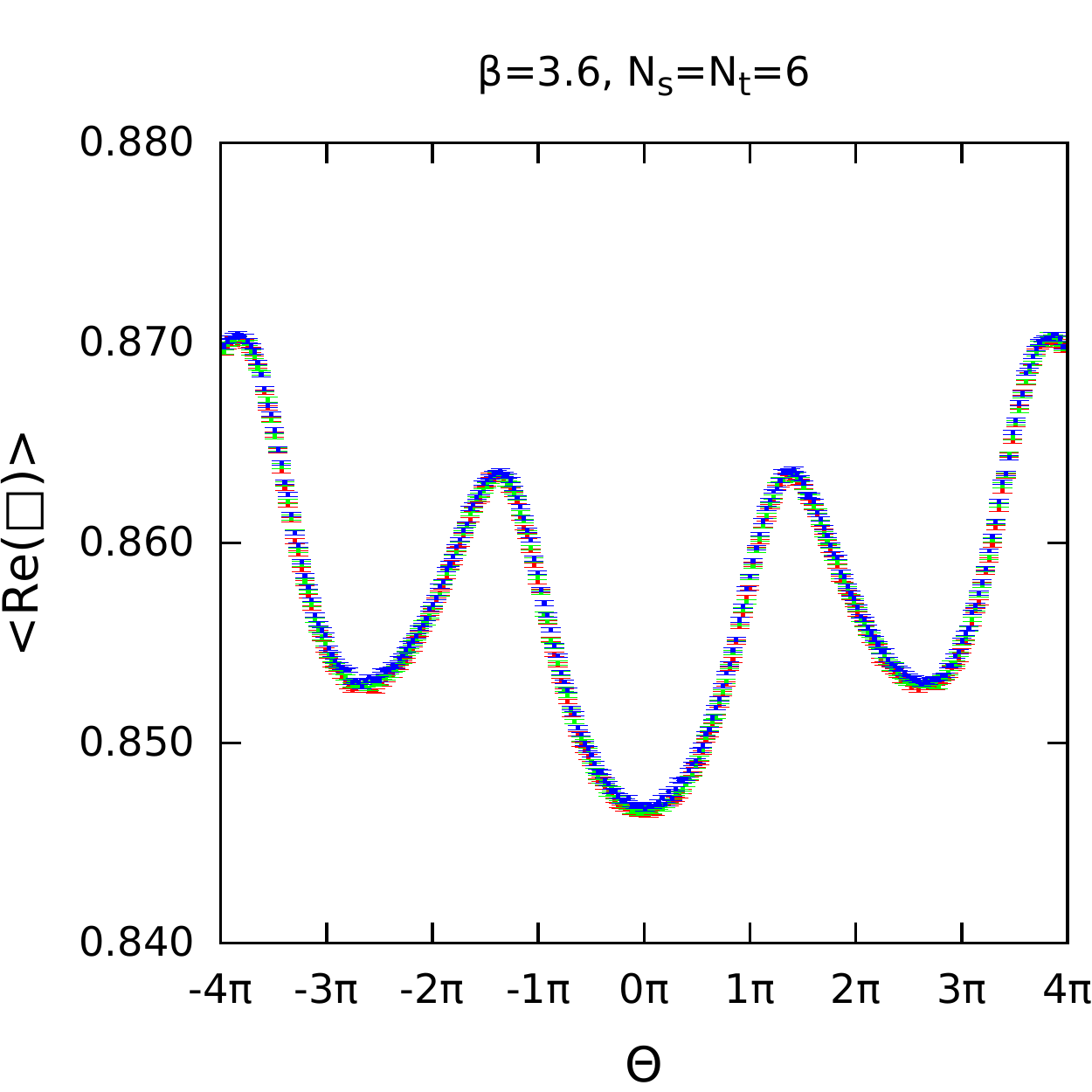}

\includegraphics[width=7.516cm,type=pdf,ext=.pdf,read=.pdf]{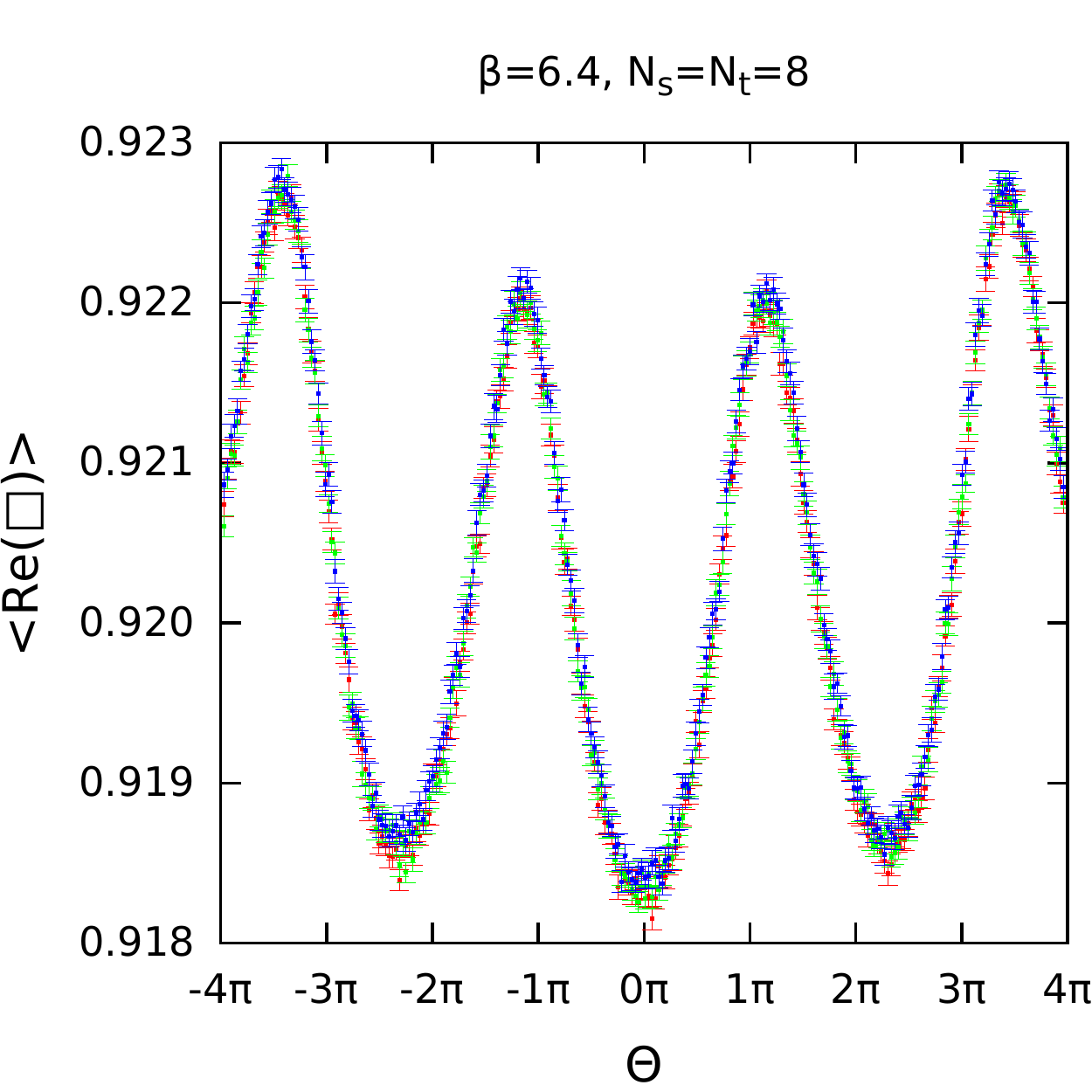}
\includegraphics[width=7.516cm,type=pdf,ext=.pdf,read=.pdf]{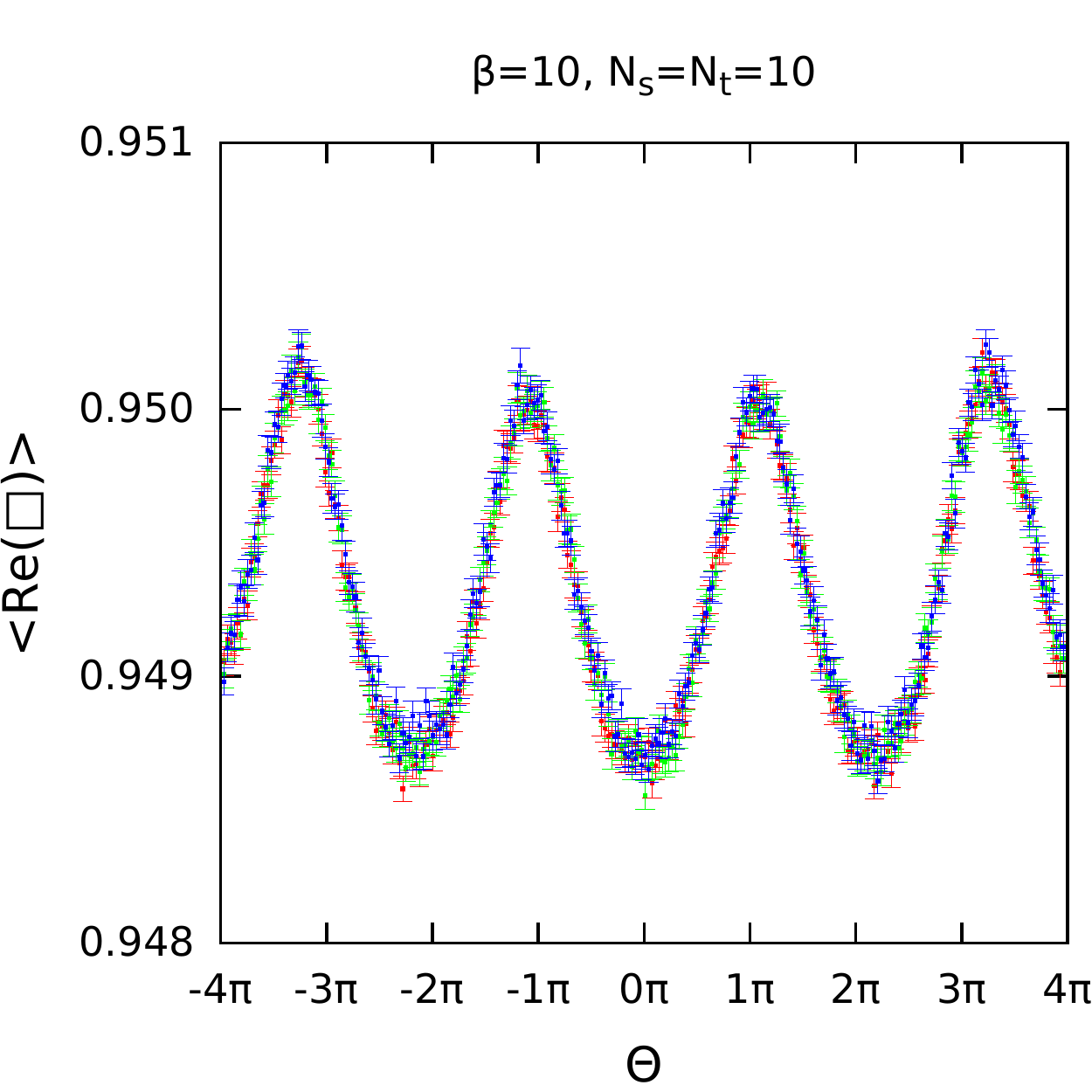}
\end{center}
\caption{Plaquette expectation value \(\langle \text{Re} (\square) \rangle\) of the full model versus the vacuum angle \(\theta\) 
for three different values of  the mass parameter \(\kappa\). 
We show the approach to the continuum limit using \(\beta = 1.6, 3.6, 6.4\) and \(10.0\) at fixed \(\beta/N_s N_t = 0.1\), using $10^6$ measurements 
for each parameter set. 
The statistical error is computed with the jackknife method.}
\label{fig_plaq}
\end{figure}

\newpage

We simulate lattices of size \(N_s=N_t=4\), \(6\), \(8\) and \(10\) at a fixed ratio \(\beta/N_s\,N_t=0.1\). Due to space restrictions in Fig. (\ref{fig_plaq}) we only show the results for the plaquette. The plots nicely demonstrate that \(\langle \text{Re} (\square) \rangle\) 
becomes \(2\pi\)-periodic in \(\theta\) when approaching the continuum limit. It is also obvious, that the plaquette shows only a very weak 
dependence on the mass parameter $\kappa$ (in the Coulomb phase) and is almost identical to the pure gauge case. The expectation value of the 
topological charge shows a similar behavior, except that it is an odd observable in \(\theta\). We can conclude that the topological charge in the dual 
formulation becomes integer valued in the combined continuum and thermodynamical limit, which is reached \textit{very quickly}, i.e., already on 
rather small lattice sizes and moderately large $\beta$.

\section{Summary}

In this contribution it was demonstrated for the first time, that a complex action problem coming from adding a topological term can be overcome
by mapping the theory to dual variables. In the dual representation the degrees of freedom are loops for the matter fields and plaquette 
occupation numbers for the gauge fields. We present a suitable update algorithm and in this exploratory study analyze bulk observables, which 
have a particularly simple form in the dual representation. We consider a suitable continuum and thermodynamical double-limit and demonstrate 
that the observables become $2\pi$-periodic in that limit, indicating that the field theoretical definition of the topological charge (which is not an 
integer at finite lattice spacing) is capable of reproducing the expected $\theta$-dependence.

\section*{Acknowledgements}
We thank Ydalia Delgado Mercado, Christian Lang, Michael M\"uller-Preu\ss ker, Hans-Peter Schadler, Alexander Schmidt and Andreas Wipf for 
discussions. This work is partly supported by DFG TR55, \textit{"Hadron Properties from Lattice QCD"} and by the Austrian Science Fund FWF 
Grant. Nr. I 1452-N27. T. K. is supported by the FWF DK W1203, \textit{"Hadrons in Vacuum, Nuclei and Stars"}.

\end{document}